\documentclass[aps,prl,twocolumn,superscriptaddress,longbibliography,nofootinbib,notitlepage]{revtex4-1}
\usepackage{amsmath}
\usepackage{amstext}
\usepackage{amssymb}
\usepackage{xfrac}
\usepackage[colorlinks,citecolor=blue]{hyperref}
\usepackage{graphicx}
\usepackage{amsmath}
\usepackage{amstext}
\usepackage{amssymb}
\usepackage{amsfonts}
\usepackage{longtable,booktabs}
\usepackage{hyperref}\usepackage{url}
\usepackage{subfigure}%
\usepackage{dsfont}
\usepackage{booktabs}
\usepackage{amsbsy}
\usepackage{dcolumn}
\usepackage{amsthm}
\usepackage{bm}
\usepackage{esint}
\usepackage{multirow}
\usepackage{hyperref}
\hypersetup{
    colorlinks=true,
    linkcolor=blue,
    filecolor=magenta,
    urlcolor=blue,
}
\usepackage{cleveref}

\usepackage{mathrsfs}
\usepackage{amsfonts}
\usepackage{amsbsy}
\usepackage{dcolumn}
\usepackage{bm}
\usepackage{multirow}
\usepackage{color}
\usepackage{ulem}
\usepackage{float}

\newcommand{\comments}[1]{}

  \usepackage{extarrows}
\usepackage{datetime}

\usepackage{comment}
\usepackage[super]{nth}

\begin{document}

\title{Quantum Hydrodynamics of Fractonic Superfluids with Lineon Condensate: from   Navier-Stokes-like Equations to   Landau-like Criterion}

\author{Jian-Keng Yuan}
\affiliation{School of Physics, State Key Laboratory of Optoelectronic Materials and
Technologies, Sun Yat-sen University, Guangzhou, 510275,
China}
\author{Shuai A. Chen}
\email{chsh@ust.hk}
\affiliation{Department of Physics, The Hong Kong University of Science and Technology, Hong Kong SAR, China}
\author{Peng Ye}
\email{yepeng5@mail.sysu.edu.cn}
\affiliation{School of Physics, State Key Laboratory of Optoelectronic Materials and
Technologies, Sun Yat-sen University, Guangzhou, 510275,
China}
 \begin{abstract}
Fractonic superfluids are exotic states of matter with spontaneously broken higher-rank $U(1)$ symmetry. The latter is associated with conserved quantities that include not only particle number (i.e. charge) but also higher moments, such as dipoles, quadrupoles, and angular moments. Due to the presence of such conserved quantities, the mobility of particles is restricted either completely or partially. In this work, we systematically study hydrodynamical properties of fractonic superfluids, especially focusing on the fractonic superfluids with conserved angular moments. The constituent bosons are called ``lineons'' with $d$-components in $d$-dimensional space. From
Euler-Lagrange equation, we derive the continuity
equation and Navier-Stokes-like equations, 
in which the angular moment conservation introduces extra terms. Furthermore, we discuss the current configurations that are related to 
the defects. Like the conventional superfluid, we study the critical
values of velocity fields and density currents, which gives rise to a   Landau-like criterion.  At the end of this work, several future directions are discussed. 
\end{abstract}
\maketitle

Superfluidity is a well-known macroscopic quantum phenomenon in which interacting bosons are organized into a macroscopic coherent state. Theoretically, we can realize the superfluid phase in a weekly interacting boson system. 
In the presence of
positive chemical potential and repulsive interaction, the $U(1)$ symmetry
is spontaneously broken and bosons are condensed.   
 Recently, an unconventional superfluid, namely, fractonic superfluid~\cite{2020PhRvR2b3267Y,Chen2021FS,Ye2021PRR,Yuan:2022mns}, has been proposed, which is motivated by  the impressive ongoing progress in the field of ``fracton topological order''~\cite{Nandkishore2019,2020Fracton}. The basic idea of fractonic superfluids is to consider quantum many-body physics of fracton-like particles. Instead of regarding fractons, lineon, and other subdimensional particles as topological excitations of fracton topological order states \cite{Chamon05fracton, Haah11, Yoshida13, Fu16,Nandkishore2019,2020Fracton,Hermele17, Vijay17coupledLayersXcube, Shirley2018, Chen20coupledCS, Aasen20tqftnetwork, Slagle21foliatedfield, LiYe20fracton, LiYe21fracton,XuWu08RVP, Pretko17tensorgauge, Chen18tensorhiggs, Barkeshli18tensorhiggs, Seiberg21ZNfracton,Pretko18fractongaugeprinciple, Gromov19multipoleAlgebra, Seiberg19vectorSym,2020PhRvR2b3267Y,Chen2021FS,Ye2021PRR,Yuan:2022mns,Zhu:2022lbx,Nandkishore21sym}, we regard them as constituent particles of real many-body systems.  In fractonic superfluids, these particles are bosonic, and the conserved quantities include not only particle number, but also higher moments, such as dipoles, quadrupoles, and angular moments~\cite{Pretko2018,PhysRevX.9.031035,Seiberg2019arXiv1909,Gorantla:2022eem}. The associated $U(1)$ symmetry is said to be higher-rank, while the usual $U(1)$ symmetry with particle number conservation is rank-$0$. Bearing this in mind, fractonic superfluids are actually symmetry-breaking phases with spontaneously broken higher-rank symmetry (HRS). Recently, this line of thinking with  general considerations of higher moment conservation and symmetry breaking has been focused on  by both condensed matter  and  high energy community~\cite{PhysRevD.104.105001,Bidussi:2021nmp,Jain:2021ibh,Angus:2021jvm,Grosvenor:2021hkn,Banerjee:2022unj,Nandkishore21sym}.

In fractonic superfluids, the constituent bosons are subject to  restriction of mobility. If the constituent bosons do not have quadratic kinetic energy (hopping term), the mobility of such bosons (fractons) is completely restricted, which can be realized by implementing dipole conservation. Likewise, if the constituent bosons have $d$ components and  the $i$-th component has a quadratic kinetic energy only along the spatial direction $\hat{x}_i$, the mobility of such bosons (lineons) is partially restricted, which can be realized by implementing angular moment conservation. 
Following the notation introduced before, we may use $d\mathsf{SF}^0$ to represent fractonic superfluid phase with fracton condensation in $d$-dimensional space, where the superscript ``$0$'' means that the constituent particles do not have kinetic energy at all. Likewise, $d\mathsf{SF}^1$ represents fractonic superfluid phase with lineon condensation in $d$-dimensional space, where the superscript ``$1$'' encodes  the above definition of lineons. In this notation, the conventional superfluid is represented by $d\mathsf{SF}^d$ in which bosons have a normal kinetic energy along all $d$ spatial directions.

As a quantum fluid, we expect fractonic superfluid should exhibit exotic quantum hydrodynamical behaviors that are entirely distinct from the conventional superfluid denoted by $d\mathsf{SF}^{d}$. In this paper, we explore  highly unconventional hydrodynamics of a fractonic superfluid with lineon condensation, i.e., $d\mathsf{SF}^1$ originally introduced in Ref.~\cite{Chen2021FS}.   
Following the Euler-Lagrange equation, we subsequently derive the continuity equation as well 
as the Navier-Stokes-like hydrodynamic equation. 
Specifically in $d=2$, we discuss the velocity fields   and the 
vorticity of superfluid defects. We find a Landau-like criterion for $d\mathsf{SF}^1$ to survive against current occurrence. 
Lastly, we make a conclusion as well as discussions on future directions.



\textit{Model and symmetry.--}
When a many-body system respects conservation law of multi-moments, 
then the constituent particles are subject to   mobility constraint. 
One example is   ``lineon'', which only moves in one certain direction in $d$-spatial dimensions, in contrast to a conventional particle that can freely
propagate in the whole space. 
To begin with, we consider a system of many $d$-component bosons with Hamiltonian \cite{Chen2021FS},
\begin{align}
\mathcal{H}= & \sum_{a=1}^{d}\partial_{a}\hat{\Phi}_{a}^{\dagger}\partial_{a}\hat{\Phi}_{a}+\sum_{a,b\neq a}^{d}\frac{1}{2}K_{ab}\left(\hat{\Phi}_{a}^{\dagger}\partial_{a}\hat{\Phi}_{b}^{\dagger}+\hat{\Phi}_{b}^{\dagger}\partial_{b}\hat{\Phi}_{a}^{\dagger}\right)\nonumber \\
 & \times\left(\hat{\Phi}_{a}\partial_{a}\hat{\Phi}_{b}+\hat{\Phi}_{b}\partial_{b}\hat{\Phi}_{a}\right)+V\left(\hat{\Phi}^{\dagger},\hat{\Phi}\right)\,,\label{eq:Hamiltonian}
\end{align}
where $\hat{\Phi}_{a}^{\dagger}$ and $\hat{\Phi}_{a}$ are respectively creation and annihilation operators with a bosonic communicative relation, 
  $\left[\hat{\Phi}_{a}(\mathbf x),\hat{\Phi}_{b}^{\dagger}(\mathbf y)\right]=\delta_{ab}\delta^{d}\left(\mathbf x-\mathbf y\right) 
 $ 
with the coordinates $\mathbf x=(x^1,\cdots, x^d)$.
The first terms in Eq.~\eqref{eq:Hamiltonian} denote the kinetic energy of $a$th component freely propagating in $a$-direction. The second term  
describes the cooperative motion involving a two components with the coefficients $K_{ab}=K_{ba}>0$ and $K_{aa}=0$. In other words, the specified form of kinetic terms impose a constraint on mobility, which is manifested by the invariant under the transformations,
$(\hat{\Phi}_{a},\hat{\Phi}_{b})\rightarrow (\hat{\Phi}_{a}e^{i\lambda_{ab}x_{b}},\hat{\Phi}_{b}e^{-i\lambda_{ab}x^{a}})$, 
aside from the particle number conservation of each component,
$\hat{\Phi}_{a}\rightarrow e^{i\lambda_{a}}\hat{\Phi}_{a}$. 
The according charges are the angular charge moments $\hat{Q}_{ab}=\int d^{d}x\left(\hat{\rho}_{a}x_{b}-\hat{\rho}_{b}x_{a}\right)$
and the particle number $\hat{Q}_{a}=\int d^{d}x\hat{\rho}_{a}$.
Here $\hat \rho_a=\hat \Phi^\dagger_a\hat \Phi_a$ is a density operator.
Due to the conservation of angular moment $\hat Q_{ab}$, a single $a$-component
becomes a lineon that can freely move only in the $a$th direction if other components have zero particle density. 
The potential $V=\sum_{a=1}^{d}\left(-\mu\hat{\rho}_{a}+\frac{g}{2}\hat{\rho}_{a}^{2}\right)$
contains chemical potential $\mu$ and short-range repulsive interaction $g>0$.

Once the chemical potential $\mu$ is tuned positive, 
the lineons get condensed to form a fractonic superfluid phase $d\mathsf{SF}^{1}$.  
The phase $d\mathsf{SF}^{1}$ is described by the order parameter, that is, the classical configuration as a plane wave, 
 $ \langle \hat{\Phi}_{a}\left(\mathbf{x}\right)\rangle=\sqrt{\rho_{0}} e^{i\theta_a^\mathrm{cl}(\mathbf x)}
  $ 
with $\rho_{0}=\mu/g$ and $\theta_{a}^{\mathrm{cl}}(\mathbf x)= \theta_{a}+\sum_{b}\beta_{ab}x^{b}$
with a finite momentum $\mathbf{k}_{a}=(\beta_{a1},...,\beta_{ad})$. Here the real parameters $\theta_a$ and $\beta_{ab}$ describe the degenerate ground state manifold with $\beta_{ab}=-\beta_{ba}$.
The low-energy theory can be obtained by an expansion around the classical configuration,
$ \hat{\Phi}_{a}(\mathbf{x})=\sqrt{\rho_{0}+\delta\rho_a(\mathbf x)}e^{i[\theta_{a}^\mathrm{cl}(\mathbf x)+\delta\theta_a(\mathbf x)]}$,
where $\delta\rho_{a}(\mathbf{x})$ and $\delta\theta_{a}(\mathbf x )$ are
density fluctuation and phase fluctuation fields, respectively.
In $d=2$ spatial dimensions, the system supports thermal defect excitations.
  Due to the particle number conservation and the angular moment conservation, one may construct three types of defects
in $2\mathsf{SF}^{1}$, 
    $\Theta_{0x}  =\ell_{0x}\left(\varphi,0\right),\quad\Theta_{0y}=\ell_{0y}\left(0,\varphi\right),  \,
\Theta_{1}  =\ell_{1}\left(\frac{y}{a}\varphi-\frac{x}{a}\ln\frac{r}{a},-\frac{x}{a}\varphi-\frac{y}{a}\ln\frac{r}{a}\right)~, 
  $ where $\varphi=\text{Arctan}\left(y/x\right)$ and $a$ is the defect
core size. Here we use the notation $\Theta\equiv (\theta_1,\theta_2)$. 
The defects $\Theta_{0x,0y}$ resemble vortices with topological charges $\ell_{0x,0y}$ in a conventional superfluid phase $2\mathsf{SF}^2$.
The defects $\Theta_1$ arise as a consequence of the angular moment conservation and the quantization of the topological charge $\ell_1$ needs a lattice for regularization \cite{Chen2021FS}.

\textit{Hydrodynamic equations.--}Ref.~\onlinecite{Chen2021FS} gives the Euler-Lagrange
equation of many lineon system in Eq.~(\ref{eq:Hamiltonian}). 
Following the Euler-Lagrange equation, 
we derive the Navier-Stokes-like equation from the perspective of hydrodynamics. 

In the framework of the coherent-state path integral, we can write down the Lagrangian 
$\mathcal L = i\sum_a \phi^*_a\partial_t \phi_a - H(\phi_a)$.
The Euler-Lagrange equation can be obtained straightforwardly:
\begin{align}
i\partial_{t}\phi_{a}= & -\partial_{a}^{2}\phi_{a}-\mu\phi_{a}+g\phi_{a}^{2}\phi_{a}^{*}\nonumber \\
 & +\sum_{b(\neq a)}K_{ab}\left(\partial_{a}\phi_{b}^{*}\right)\left(\phi_{b}\partial_{b}\phi_{a}+\phi_{a}\partial_{a}\phi_{b}\right)\nonumber \\
 & -\sum_{b(\neq a)}K_{ab}\partial_{b}\left[\phi_{b}^{*}\left(\phi_{b}\partial_{b}\phi_{a}+\phi_{a}\partial_{a}\phi_{b}\right)\right]~.
 \label{eq:EL}
\end{align}
   Furthermore, in terms of the relation $\phi_{a}=\sqrt{\rho_{a}}e^{i\theta_{a}}$,
we can reformulate the Euler-Lagrange equation \eqref{eq:EL} with respective to 
the density field $\rho_a(\mathbf x)$ and phase field $\theta_a(\mathbf x)$,
  \begin{align}
\partial_{t}\rho_{a}= & -\partial_{a}\left(2\rho_{a}\partial_{a}\theta_{a}\right)\nonumber \\
 & -\sum_{b(\neq a)}\partial_{b}\left[2K_{ab}\rho_{a}\rho_{b}\left(\partial_{a}\theta_{b}+\partial_{b}\theta_{a}\right)\right]~,\label{eq:rho_dt}\\
\partial_{t}\theta_{a}= & -\left(\partial_{a}\theta_{a}\right)^{2}+p_{a}(\rho)/2\nonumber \\
 & -\sum_{b(\neq a)}K_{ab}\rho_{b}\left(\partial_{a}\theta_{b}+\partial_{b}\theta_{a}\right)^{2}~,\label{eq:theta_dt}
\end{align}
where $p_{a}(\rho)$ is independent of the $\theta_{a}$ fields, 
\begin{align}
p_{a}(\rho)= & \sum_{b(\neq a)}\frac{K_{ab}}{2}\Bigg[-\rho_{b}\frac{\left(\partial_{b}\rho_{a}\right)^{2}}{\rho_{a}^{2}}-\frac{\left(\partial_{a}\rho_{b}\right)^{2}}{\rho_{b}}+\frac{2\partial_{b}\left(\rho_{b}\partial_{b}\rho_{a}\right)}{\rho_{a}}\nonumber \\
 & +2\partial_{a}\partial_{b}\rho_{b}\Bigg]+\frac{2\partial_{a}^{2}\sqrt{\rho_{a}}}{\sqrt{\rho_{a}}}+2\left(\mu-g\rho_{a}\right)~.\label{eq:quantum pressure}
\end{align}
In fact, Eq.~\eqref{eq:rho_dt} can be reformulated in the standard form of the continuity equation: 
\begin{equation}
\partial_{t}\rho_{a}=-\sum_{i}\partial_{i}\left(\rho_{a}v_{ai}\right)\,,
\end{equation}
where $v_{ai}$ is  the $i$-th component of the velocity  of the $a$-th component of lineon field $\phi_a$:
\begin{equation}
v_{ai}=2\partial_{a}\theta_{a}\delta_{ai}+2K_{ai}\rho_{i}\left(\partial_{i}\theta_{a}+\partial_{a}\theta_{i}\right)\left(1-\delta_{ia}\right)~.
\label{eq:velocityfield}
\end{equation}
This expression indicates a strong anisotropic velocity field  $v_{ai}$ of lineon field $\phi_a$. More explicitly:
\begin{align}
v_{ai}&=2\partial_{a}\theta_{a}\,(\text{if }a=i),\\
v_{ai}&=2K_{ai}\rho_{i}\left(\partial_{i}\theta_{a}+\partial_{a}\theta_{i}\right)\, (\text{if }a\neq i).
\label{eq:velocityfield111}
\end{align}
Therefore, for the $a$-th component of lineon field, i.e., $\phi_a$, there is always a nonvanishing velocity along $a$-th spatial direction if the phase field $\theta_a$ has nonvanishing gradient along $a$-th spatial direction.  This is similar to the usual velocity field of conventional superfluid. But when $i \neq a$, in order to have nonzero $v_{ai}$, i.e., nonzero velocity along $i$-th spatial direction, we at least require that $\rho_i$ is nonzero. Consequently, when $i\neq a$, $\phi_a$ and $\phi_i$ must be combined together such that $\phi_a$ can gain velocity along $i$-th direction. In other words, a single lineon $\phi_a$ can only move along $a$-th direction; $\phi_a$ may move along other directions only in the form of collective motion. An alternative explanation from symmetry argument is given in Appendix A of Ref.~\cite{Ye2021PRR}. 
Keeping this scenario in mind, for $\phi_a$, we denote $a$-direction [cf. $i=a$ in Eq.~\eqref{eq:velocityfield}] as movable direction and others [cf. $i\neq a$ in Eq.~\eqref{eq:velocityfield}] as immovable direction.
On the other hand, from Eq.~(\ref{eq:theta_dt}), by means of the velocity fields in Eq.~\eqref{eq:velocityfield},
we can obtain the following \textit{Navier-Stokes-like equation}
\begin{align}
\partial_{t}v_{ai}= & \delta_{ai}\left(-\partial_{i}\frac{v_{aa}^{2}}{2}-\sum_{b\neq a}\frac{\partial_{i}v_{ab}^{2}}{2K_{ab}}+\partial_{a}p_{a}\right)\nonumber \\
 & +\left(1-\delta_{ai}\right)\Bigg[-\sum_{b\neq a}K_{ai}\rho_{i}\left(\frac{\partial_{i}v_{ab}^{2}}{2K_{ab}}+\frac{\partial_{a}v_{ib}^{2}}{2K_{ib}}\right)\nonumber \\
 & -K_{ai}\rho_{i}\frac{\partial_{i}v_{aa}^{2}+\partial_{a}v_{ii}^{2}}{2}-\frac{v_{ai}}{\rho_{i}}\sum_{b}\partial_{b}\left(\rho_{i}v_{ib}\right)\nonumber \\
 & +K_{ai}\rho_{i}\left(\partial_{i}p_{a}+\partial_{a}p_{i}\right)\Bigg]~.
 \label{eq:Navier-Stokes eq}
\end{align}
We introduce the notations  $T_a$ of $a$-component:
\begin{equation}
T_{a}=\frac{v_{aa}^{2}}{2}+\sum_{b(\neq a)}\frac{v_{ab}^{2}}{2K_{ab}\rho_{b}}\label{eq:kinetic density}~.
\end{equation}
Then the Navier-Stokes-like equations for the diagonal components $v_{aa}$ ($a=1,\cdots,d$) in Eq.~(\ref{eq:Navier-Stokes eq}) can be written in a more compact form:
\begin{equation}
\partial_{t}v_{aa}=\partial_{a}(-T_{a}+p_{a})~.
\label{eq:NS_diagonal}
\end{equation}
For the off-diagonal components $v_{ai}$ ($i\neq a$), we have
\begin{align}
\partial_{t}v_{ai}= & K_{ai}\rho_{i}[\partial_{i}(-T_{a}+p_{a})+\partial_{a}(-T_{i}+p_{i})]\nonumber \\
 & -\frac{v_{ai}}{\rho_{i}}\sum_{b}\partial_{b}(\rho_{i}v_{ib})~.
 \label{eq:NS_offdiagonal}
\end{align}
The diagonal components in Eq.~(\ref{eq:NS_diagonal}) is similar to
the Navier-Stokes equation in conventional superfluid $d\mathsf{SF}^{d}$ \cite{rogel2013gross},
\begin{align}
\partial_{t}v_{i}=\partial_{i}\left(-\sum_{j}\frac{v_{j}^{2}}{2}+\frac{1}{2\sqrt{\rho}}\sum_{j}\partial_{j}^{2}\sqrt{\rho}-g\rho\right)\,,\label{eq:NS_conventional}
\end{align}
where $v_i$ and $\rho$ are the velocity fields and density field in $d\mathsf{SF}^{d}$, respectively.
Here,  $\sum_{j}v_{j}^{2}/2$ is the kinetic density and $\frac{1}{2\sqrt{\rho}}\sum_{j}\partial_{j}^{2}\sqrt{\rho}$ is the quantum pressure term in $d\mathsf{SF}^{d}$.
The comparison with Navier-Stokes equation in $d\mathsf{SF}^{d}$ gives the meaning
of $T_{a}$ in Eq.~(\ref{eq:kinetic density}) as kinetic density and
$p_{a}$ in Eq.~(\ref{eq:quantum pressure}) as pressure term from
lineon effects. The kinetic density of immovable direction $x^{b}$
contains the $b$-component density field $\rho_{b}(\mathbf{x})$,
may characterize the difference between fractonic hydrodynamics
and conventional hydrodynamics. 

For the off-diagonal components in Eq.~(\ref{eq:NS_offdiagonal}), the density $\rho_{i}\left(\mathbf{x}\right)$ outside the partial
derivative makes currents in the immovable direction beyond the conventional
Navier-Stokes equation.
However, for the ground state of $d\mathsf{SF}^{1}$, i.e. $\rho_{0}=\mu/g$ and $\theta_{a}^{\mathrm{cl}}$, 
the pressure term $p_{a}\equiv0$ and the off-diagonal components in Eq.~\eqref{eq:NS_offdiagonal}
can be further simplified:
\begin{equation}
\partial_{t}v_{ai}=-K_{ai}\rho_{0}(\partial_{i}T_{a}+\partial_{a}T_{i})-v_{ai}\sum_{b}\partial_{b}v_{ib}~,
\label{eq:NS off diagonal condensed}
\end{equation}
where the first term is similar to those in Eq.~(\ref{eq:NS_conventional}) and 
the second term is the pressure term from the $i$-component
to the $a$-component. The physical meaning of the third
term in Eq.~(\ref{eq:NS off diagonal condensed}) is still an open
question.

In addition to the above discussion, we can also study the velocity fields in the presence of topological defects caused by the multi-valuedness of phase fluctuation field $\theta\left(\mathbf{x}\right)$,
playing an important role in Kosterlitz-Thouless transition of superfluids.
For notation simplification, we still use $\theta(\mathbf{x})$ to denote the phase fluctuations.
The formation of the bound states will eliminate the divergent energy
of a single defect. In $2\mathsf{SF}^{1}$, the bound states of $\Theta_{0x,0y}$
constitute with two defects when the bound states of $\Theta_{1}$
constitute with 4 defects \cite{Yuan:2022mns}. 

\begin{figure}[t]
\centering{}\includegraphics[scale=1.212]{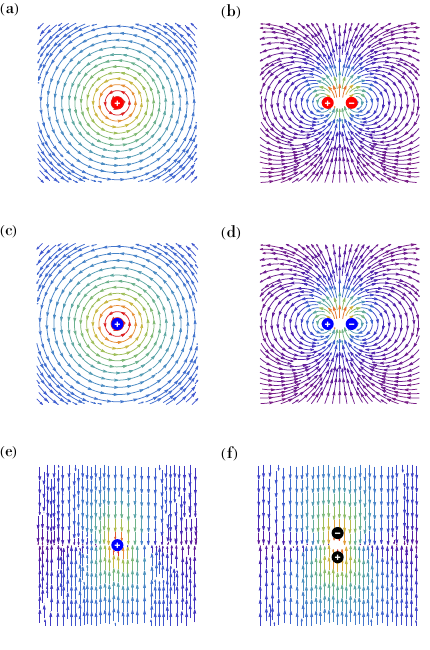} 
\caption{The velocity fields in Eq.~(\ref{eq:velocityfield}) of defects in $2\mathsf{SF}^1$: 
$(v_{11},v_{12})$ in
(a, b, e) and  $(v_{12},v_{22})$ in (c, d, f). 
 The color and directions of arrows represent the strength (the brighter
color the larger strength) and direction of the velocity fields. The
red, blue and black dot mark the cores of the defects $\Theta_{0x},\Theta_{0y}$
  and $\Theta_{1}$  
with topological charge $\pm1$, respectively. 
\label{fig:multidefects}}
\end{figure}

In two spatial dimensions, we have four components velocity fields $v_{ai}$ ($a,i=1,2$). In $2\mathsf{SF}^1$, we discuss the vorticity to the topological charges of defects. For a single defect $\Theta_{0x}$ with the form with $\theta_2=0$,
 we have the velocity fields ($K_{12}=K,\rho_0=\mu/g$)
\begin{equation}
v_{11}= 2\partial_1\theta_1, v_{12}= 2K\rho_{0}\partial_2\theta_1,v_{22}=0~.
\end{equation}
Then, the topological charge can be obtained from the vorticity
\begin{equation}
\ell_{0x}= \oint\left(\frac{v_{11}}{2}dx +\frac{v_{12}}{2K\rho_{0}}dy\right)~.
\end{equation}
It resembles a conventional vortex in $2\mathsf{SF}^2$. Similarly,
for a single $\Theta_{0y}$ with $\theta_1=0$, the velocity fields take the form as 
\begin{equation}
v_{11}=0 , v_{12}= 2K\rho_{0}\partial_1\theta_2,v_{22}=2\partial_2\theta_2~.
\end{equation}
and we have the vorticity
\begin{equation}
\ell_{0y}= \oint \left(\frac{v_{12}}{2K\rho_{0}}dx +\frac{v_{22}}{2}dy\right)~.
\end{equation}
In Fig.~\ref{fig:multidefects}(a)-(d), we depict the velocity fields $(v_{11},v_{12})$ and $(v_{12},v_{22})$
for the $\Theta_{0x}$ and $\Theta_{0y}$, respectively.

The defect $\Theta_{1}$ has no correspondence in a conventional superfluid
phase. As pointed out in Ref.~\onlinecite{Chen2021FS}, a bound state
of two $\Theta_{1}$ with opposite charges reduces to $\Theta_{0x}$
and $\Theta_{0y}$, which reflects on their corresponding velocity
fields. The velocity fields $\left(v_{21},v_{22}\right)$ of two $\Theta_{1}$
with opposite charges are similar to the velocity fields $\left(v_{11},v_{12}\right)$
of a single $\Theta_{0y}$ [Figs.~\ref{fig:multidefects}(e) and~\ref{fig:multidefects}(f)], and velocity
fields $\left(v_{11},v_{12}\right)$ of two $\Theta_{1}$ with opposite
charges to $\left(v_{21},v_{22}\right)$ of a single $\Theta_{0x}$. 

\textit{Landau-like criterion and critical currents.--}
In the lineon condensation phase $d\mathsf{SF}^1$, the occurrence of currents 
 does harm to the superfluidity. 
Here, we discuss the critical currents which is the critical value for the stability of $d\mathsf{SF}^1$.

For simplicity, we consider the isotropic case $K_{ab}=K$ with the uniform velocity fields. In presence of uniform
superfluid density $\rho_{a}(\mathbf{x})=\rho$, 
one may define 
 $ \nu=\sum_{a}\left|\partial_{a}\theta_{a}\right|^{2}/d\,, w=\sum_a\sum_{b(\neq a)}\left|\partial_{a}\theta_{b}+\partial_{b}\theta_{a}\right|^{2}/d~,
$ 
 as two different average phase fluctuations.
And the energy density of $d\mathsf{SF}^{1}$ should be 
\begin{equation}
\mathcal{E}=d\left[\frac{\rho^{2}}{2}\left(g+Kw\right)+\rho\left(\nu-\mu\right)\right]~,
\end{equation}
which reaches its minimum when
\begin{equation}
\rho_{0}=\frac{\mu-\nu}{Kw+g}\label{eq: superfluid density}~.
\end{equation}
Since $\rho_{0} > 0$ in the superfluid phase, we obtain a condition $\nu <\mu$.
It implies the critical value for velocity fields in
the movable direction, $v_{\text{m}}^{\text{max}}=2\sqrt{\mu}$, which plays a role as 
Landau criterion in a conventional superfluid phase, $d\mathsf{SF}^{d}$. Furthermore, $w\equiv0$ in $d\mathsf{SF}^{d}$, a conventional superfluid phase,
as the condensed bosons can move in the whole space and $\nu\equiv0$
in $d\mathsf{SF}^{0}$ as a single fracton is totally immovable.
Here in $d\mathsf{SF}^{1}$, the current
\begin{align}
J_{\text{m}}=\sum_{a}\left|J_{aa}\right|/d
\end{align}
 is along the movable direction and the current 
 \begin{align}
 J_{\text{im}}=\sqrt{\sum_{i}\sum_{a(\neq i)}\left|J_{ai}\right|^{2}/d}
 \end{align}
 is along the immovable direction. These two currents have different critical values. For $\nu=\mu/3,w=0$, currents only occur in the movable direction and $J_\mathrm{m}$ becomes maximal with 
\begin{equation}
J_{\text{m}}^{\text{max}}  =\frac{4\mu^{3/2}}{3\sqrt{3}g}\label{eq:m critical current}~,
\end{equation}
while for $\nu=0,w=g/3K$, currents $J_\mathrm{im}$ only occur in the
immovable direction with the maximal value
$J_{\text{im}}^{\text{max}}$
\begin{align}
J_{\text{im}}^{\text{max}} & =\frac{3\sqrt{3K}\mu^{2}}{8g^{3/2}}~.
\label{eq:im critical current}
\end{align}
Once the currents exceeds the critical values, the superfluid density vanishes
and the system is no longer in the superfluid phase. In addition, we can also expect thermal fluctuations may destroy superfluid phase, which has been studied in Refs.~\cite{Chen2021FS,Yuan:2022mns}.

\textit{Outlook.--}In this paper, we derive the continuity equations and Navier-Stokes-like
equations of a many-lineon system. We study the critical currents of
fractonic superfluid phase $d\mathsf{SF}^{1}$ after   lineons are condensed.
The currents in the movable direction behave like currents in a conventional superfluid denoted by $d\mathsf{SF}^{d}$
while the currents in immovable direction behave like currents in
$d\mathsf{SF}^{0}$. 
With the equations established in this work, in the future
work, we can study the basic hydrodynamic properties of $d\mathsf{SF}^{1}$,
like compressibility, viscosity and irrotationality.  As the equations are written with general $d$, it will be interesting to investigate the potential dimension reduction and underlying physical consequences \cite{lapa17}. By putting $d\mathsf{SF}^{1}$ on a curved space, one can further study how background gravitational field enters into the Navier-Stokes-like equations and how hydrodynamical field is entangled with geometric quantities of based manifold~\cite{Bidussi:2021nmp,PhysRevB.99.205120}. Finally, we can also regard bosonic lineon field $\Phi$ as the vacuum expectation of a bilinear form of a more fundamental fermion such that the Hamiltonian we wrote is a mean-field theory of a more fundamental fermion system in the language of projective construction by using techniques in, e.g., \cite{YW13a,YW12,ye16a}. Along this line of thinking, it is interesting to ask the analytic change of hydrodynamical properties in the presence of gauge fluctations, which potentially signals exotic non-Fermi liquid theory or spin liquid states.  
We can also regard hydrodynamical behaviors in open systems, where an exotic non-Hermitian quantum effect may be expected  \cite{ashidaNonHermitian2020} and nontrivial entanglement properties \cite{chenEntanglement2020,leePositionmomentum2014} may be hidden in a set of non-Hermitian hydrodynamical equations.

\textit{Acknowledgement.--}The work was supported by Guangdong Basic and Applied Basic Research Foundation under Grant No.~2020B1515120100, NSFC Grant (No.~11847608 \& No.~12074438).

\end{document}